\documentclass[10pt,letterpaper]{article}
\usepackage{opex2}

\usepackage[dvips]{epsfig}
\usepackage{float}
\usepackage{amsmath}
\usepackage{amssymb}
\usepackage{enumerate}
\usepackage{hhline}
\usepackage{supertabular}
\usepackage{pslatex}
\usepackage{multirow}
\usepackage{array}
\usepackage{tabularx}

\newcommand{\frakb}[1]{\frak{#1}}
\newcommand{\vecb}[1]{\mathbf{#1}}

\newcolumntype{Y}{>{\centering\arraybackslash}X}

\begin{document}

\title{Cavity $Q$, mode volume, and lasing threshold in small diameter AlGaAs microdisks with embedded quantum dots}

\author{Kartik Srinivasan, Matthew Borselli, and Oskar Painter}
\address{Department of Applied Physics, California Institute of Technology, Pasadena, CA 91125, USA.}
\email{phone: (626) 395-6269, fax: (626) 795-7258, e-mail: kartik@caltech.edu}
\author{Andreas Stintz and Sanjay Krishna}
\address{Center for High Technology Materials, University of New Mexico, Albuquerque, NM 87106, USA.}

\begin{abstract}
The quality factor ($Q$), mode volume ($V_{\text{eff}}$), and room-temperature lasing threshold of microdisk cavities with embedded quantum dots (QDs) are investigated.  Finite element method simulations of standing wave modes within the microdisk reveal that $V_{\text{eff}}$ can be as small as 2$(\lambda/n)^3$ while maintaining radiation-limited $Q$s in excess of 10$^5$.  Microdisks of diameter 2 $\mu$m are fabricated in an AlGaAs material containing a single layer of InAs QDs with peak emission at $\lambda=1317$ nm. For devices with $V_{\text{eff}}\sim$2 $(\lambda/n)^3$, $Q$s as high as $1.2{\times}10^5$ are measured passively in the 1.4 $\mu$m band, using an optical fiber taper waveguide.  Optical pumping yields laser emission in the $1.3$ $\mu$m band, with room temperature, continuous-wave thresholds as low as $1$ $\mu$W of absorbed pump power.  Out-coupling of the laser emission is also shown to be significantly enhanced through the use of optical fiber tapers, with laser differential efficiency as high as $\xi\sim16\%$ and out-coupling efficiency in excess of $28\%$.
\end{abstract}


\setcounter{page}{1}
\noindent
\section{Introduction}
\label{sec:intro}
Optical microcavities with embedded quantum dots (QDs) have become a very active area of research, with applications to triggered single photon sources\cite{ref:Michler,ref:Santori,ref:Moreau}, strongly coupled light-matter systems for quantum networking\cite{ref:Reithmaier,ref:Yoshie3,ref:Peter}, and low threshold microcavity lasers\cite{ref:Cao,ref:Ide2,ref:Yang_T2}.  For these applications some of the most important microcavity parameters are the quality factor ($Q$), mode volume ($V_{\text{eff}}$), and the efficiency of light collection from the microcavity ($\eta_{o}$).  $Q$ and $V_{\text{eff}}$ describe the decay rate ($\kappa$) and peak electric field strength within the cavity, respectively, which along with the oscillator strength and dephasing rate of the QD exciton determine if the coupled QD-photon system is in the regime of reversible energy exchange (strong coupling) or in a perturbative regime (weak coupling) characterized by a modification of the QD exciton radiative lifetime (the Purcell effect)\cite{ref:Kimble2}.  The collection efficiency $\eta_{0}$ is of great importance for quantum networking\cite{ref:Cirac} and linear optics quantum computing applications\cite{ref:Knill,ref:Kiraz}, where near-unity photon pulse collection values are required.        

Microdisks supporting high-$Q$ whispering-gallery resonances were first studied in the context of semiconductor microlasers in the early 1990s\cite{ref:McCall2}.  Since that time there has been extensive work on incorporating self-assembled InAs QD active regions within semiconductor microdisks for studying quantum interactions of light and matter\cite{ref:Michler,ref:Peter,ref:Cao,ref:Ide2,ref:Yang_T2,ref:Gayral}.  With respect to lasers, the relatively small modal gain available from a single layer of QDs has typically resulted in device operation at reduced temperatures\cite{ref:Cao}, or the use of multiple QD layers to achieve room temperature (RT) operation\cite{ref:Ide2,ref:Yang_T2}.  More recently, improvements in the cavity $Q$ have resulted in RT operation in devices containing a single layer of QDs\cite{ref:Srinivasan9}.  Furthermore, it has been shown that the collection efficiency of emitted power can be significantly increased by using optical fiber tapers to evanescently couple light from the microdisk\cite{ref:Srinivasan11}.  As discussed above, these improvements in $Q$ and $\eta_{0}$ are not only important for lasers, but for future experiments in cavity quantum electrodynamics (cQED).    

In this article, we continue our study of taper-coupled microdisk-QD structures by considering device performance as the disks are scaled down in size.  In Section {\ref{sec:sims}}, we use finite element simulations to examine the behavior of $Q$ and $V_{\text{eff}}$ as a function of disk diameter.  We relate these parameters to those used in cQED, and from this, determine that disks of $1.5-2$ $\mu$m in diameter are optimal for use in future experiments with InAs QDs.  Section {\ref{sec:setup}} briefly outlines the methods used to fabricate and test devices consisting of a 2 $\mu$m diameter disk created in an AlGaAs heterostructure with a single layer of self-assembled InAs QDs. In Section {\ref{sec:results}}, we present experimental measurements of the fabricated devices.  Through passive characterization, cavity $Q$s as high as $1.2{\times}10^5$ are demonstrated for devices with a predicted $V_{\text{eff}}\sim2.2(\lambda/n)^3$.  In addition, photoluminescence measurements show that the devices operate as lasers with RT, continuous-wave thresholds of $\sim$1 $\mu$W of absorbed pump power.  Finally, the optical fiber taper is used to increase the efficiency of out-coupling by nearly two orders of magnitude, so that an overall fiber-coupled laser differential efficiency of $\xi\sim16\%$ is achieved.  We conclude by presenting some estimates of the number of QDs contributing to lasing and the spontaneous emission coupling factor ($\beta$) of the devices.

\section{Simulations}
\label{sec:sims}

To study $Q$ and $V_{\text{eff}}$ of the microdisk cavities, finite-element eigenfrequency simulations\cite{ref:Spillane3,ref:Borselli3} are performed using the Comsol FEMLAB commerical software.  By assuming azimuthal symmetry of the disk structures, only a two-dimensional cross-section of the disk is simulated, albeit using a full-vectorial model.  The cavity mode effective volume is calculated according to the formula\cite{ref:Andreani}: 

\begin{equation}
\label{eq:mode_volume}
\begin{split}
V_{\text{eff}}=\frac{\int_{V} \epsilon(\vecb{r})|\vecb{E(\vecb{r})}|^2d^{3}\vecb{r}}{\max[\epsilon(\vecb{r})|\vecb{E(\vecb{r})}|^2]}\\
\end{split}
\end{equation}      

\noindent where $\epsilon(\vecb{r})$ is the dielectric constant, $|E(\vecb{r})|$ is the electric field strength, and $V$ is a quantization volume encompassing the resonator and with a boundary in the radiation zone of the cavity mode under study.  The resonance wavelength $\lambda_{0}$ and radiation limited quality factor $Q_{\text{rad}}$ are determined from the complex eigenvalue (wavenumber) of the resonant cavity mode, $k$, obtained by the finite-element solver, with $\lambda_{0}=2\pi/\frakb{Re}(k)$ and $Q_{\text{rad}}=\frakb{Re}(k)/(2\frakb{Im}(k))$.  

\begin{figure}[ht]
\begin{center}
\epsfig{figure=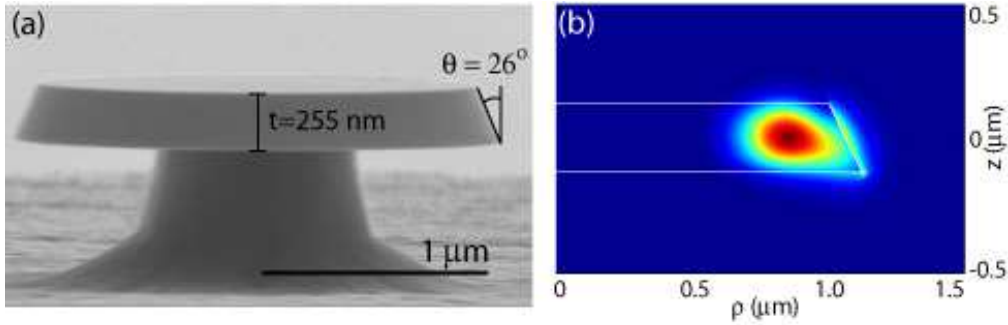, width=\linewidth}
\caption{(a) Scanning electron microscope (SEM) image of a fabricated microdisk device. The disk thickness $t$=255 nm and sidewall angle $\theta=26^{\circ}$ from vertical are taken as fixed in the finite-element simulations presented in the work. The measured average diameter for this device (i.e., the diameter at the center of the slab) is $\sim 2.12$ $\mu$m. (b) Finite-element-calculated $|\vecb{E}|^2$ distribution for the TE$_{p=1,m=11}$ WGM of a microdisk with a diameter of $\sim$ 2.12 $\mu$m at the center of the slab.  For this mode, $\lambda\sim1265.41$ nm, $Q_{\text{rad}}\sim10^7$, and $V_{\text{eff}}\sim2.8 (\lambda/n)^3$.}
\label{fig:SEM}
\end{center}
\end{figure} 

Figure \ref{fig:SEM}(a) shows a scanning electron microscope (SEM) image of a fabricated microdisk.  The devices are formed from a GaAs/AlGaAs waveguide layer that is $255$ nm thick, and due to an emphasis of sidewall smoothness over verticality during fabrication\cite{ref:Srinivasan9}, the etched sidewall angle is approximately $26^{\circ}$ from vertical.  These parameters are included in the simulations as shown Figure \ref{fig:SEM}(b).  Here, we will focus on resonant modes in the 1200 nm wavelength band, corresponding to the low temperature (T=4 K) ground-state exciton transition of the QDs, relevant for future cQED experiments.  We confine our attention to the more localized transverse electric (TE) polarized modes of the microdisk, and only consider the first order radial modes. In what follows we use the notation TE$_{p,m}$ to label whispering-gallery-modes (WGMs) with electric field polarization dominantly in the plane of the microdisk, radial order $p$, and azimuthal mode number $m$.  The refractive index of the microdisk waveguide is taken as $n=3.36$ in the simulations, corresponding to the average of the refractive indices of the GaAs and AlGaAs layers at $\lambda=1200$ nm.     

In addition, the modes that we study are \emph{standing wave} modes that are superpositions of the standard clockwise (CW) and counterclockwise (CCW) \emph{traveling wave} modes typically studied in microdisks.  These standing wave modes form when surface scattering couples and splits the initially degenerate CW and CCW traveling wave modes\cite{ref:Weiss,ref:Kippenberg,ref:Borselli2}; this process is contingent upon the loss rates within the disk being small enough for coherent coupling between the traveling wave modes to occur.  For the AlGaAs microdisk devices we have studied thus far\cite{ref:Srinivasan9,ref:Srinivasan11}, this has indeed been the case.  The effective mode volume for a standing wave mode, as defined in Equation \ref{eq:mode_volume}, is roughly half that of a traveling wave mode\cite{ref:Borselli2}.  This is of particular relevance to cQED experiments involving such microdisks as the coherent coupling rate of light and matter scales as $g\sim 1/\sqrt{V_{\text{eff}}}$.  A QD positioned at an anti-node of the standing wave will have an exciton-photon coupling rate which is $\sqrt{2}$ times larger than for the traveling wave mode.   

\begin{figure}[ht]
\begin{center}
\epsfig{figure=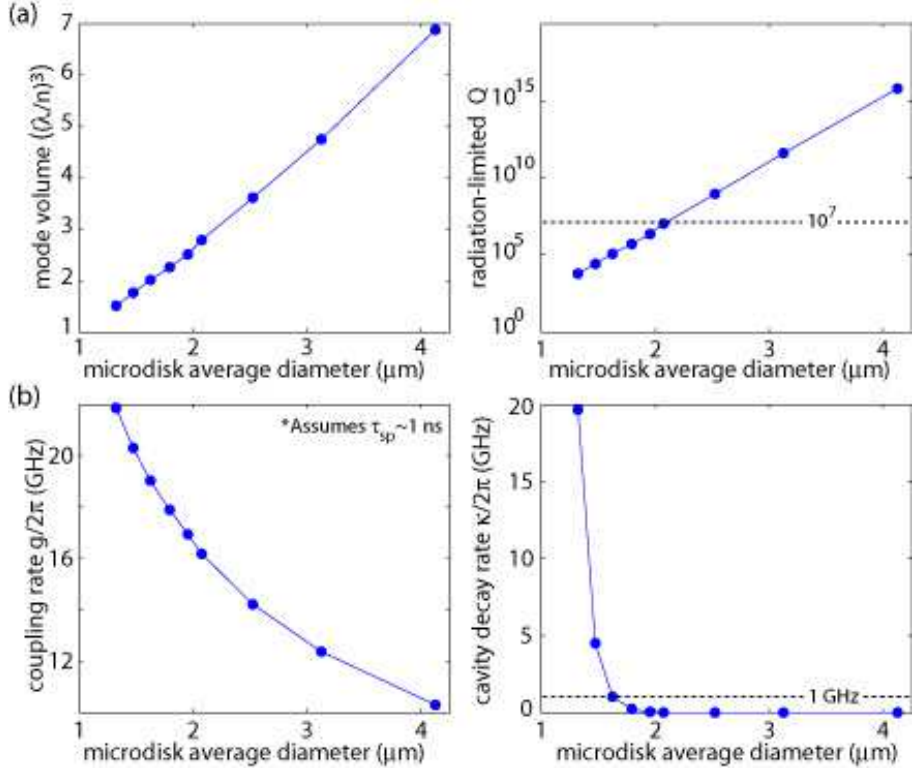, width=0.9\linewidth}
\caption{Finite-element method simulation results: (a) Modal volume $V_{\text{eff}}$ (left) and radiation-limited cavity quality factor $Q_{\text{rad}}$ (right) as a function of microdisk diameter (taken at the center of the slab), calculated for standing wave modes of disks of the shape shown in Fig. \ref{fig:SEM}.  The modes studied are TE$_{p=1,m}$ WGMs with resonance wavelength within the $1200$ nm band. (b) Coherent coupling rate $g/2\pi$ (left) and cavity decay rate $\kappa/2\pi$ (right) as a function of microdisk diameter.  A QD spontaneous emission lifetime $\tau_{sp}=1$ ns is assumed in the calculation of $g$.}
\label{fig:sim_results}
\end{center}
\end{figure} 

\begin{figure}[ht]
\begin{center}
\epsfig{figure=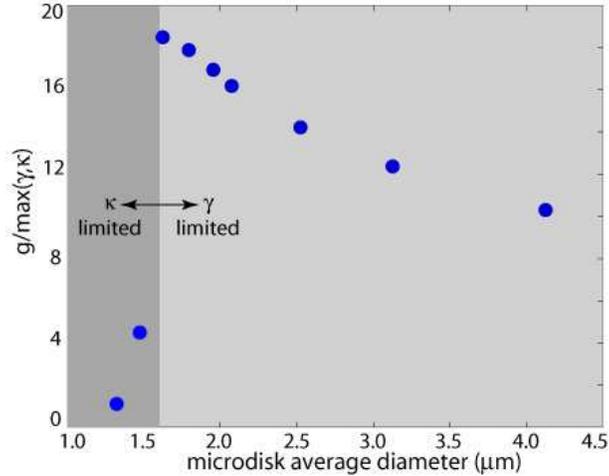, width=0.6\linewidth}
\caption{Ratio of the calculated coupling rate $g$ to the maximum decay rate in the system, max($\gamma$,$\kappa$), as a function of the microdisk diameter at the center of the slab.  A fixed QD decay rate $\gamma/2\pi$=1 GHz is assumed, and the cavity decay rate $\kappa$ is taken to be solely due to radiation loss.}
\label{fig:sim_results_2}
\end{center}
\end{figure}

Figures \ref{fig:SEM}(b) and \ref{fig:sim_results}(a) show the results of the finite element simulations.  We see that $V_{\text{eff}}$ for these standing wave modes can be as small as $2(\lambda/n)^3$ while maintaining $Q_{\text{rad}}>10^5$.  Indeed, for microdisk average diameters $D>2$ $\mu$m\footnote{The average diameter is taken at the center of the slab, or equivalently, is the average of the top and bottom diameters.}, radiation losses are not expected to be the dominant loss mechanism as $Q_{\text{rad}}$ quickly exceeds $10^7$, and other sources of field decay such as material absorption or surface scattering are likely to dominate .  To translate these results into the standard parameters studied in cQED, we calculate the cavity decay rate $\kappa/2\pi=\omega/(4{\pi}Q)$ (assuming $Q=Q_{\text{rad}}$) and the coherent coupling rate $g$ between the cavity mode and a single QD exciton.  In this calculation, a spontaneous emission lifetime $\tau_{sp}=1$ ns is assumed for the QD exciton, and $g=\vecb{d}\cdot\vecb{E}/\hbar$ is the vacuum coherent coupling rate between cavity mode and QD exciton, given by\cite{ref:Kimble2,ref:Andreani}:    

\begin{equation}
\label{eq:coupling_rate}
\begin{split}
g/2\pi=\frac{1}{2\tau_{sp}}\sqrt{\frac{3c\lambda_{0}^2\tau_{sp}}{2\pi{n^3}V_{\text{eff}}}}, \\
\end{split}
\end{equation}    

\noindent where $c$ is the speed of light and $n$ is the refractive index at the location of the QD.  This formula assumes that the QD is optimally positioned within the cavity field, so that the calculated $g$ is the maximum possible coupling rate.  
The resulting values for $g$ and $\kappa$ are displayed in Figure \ref{fig:sim_results}(b), and show that $g/2\pi$ can exceed $\kappa/2\pi$ by over an order of magnitude for a range of disk diameters.  In addition, for all but the smallest-sized microdisks, $\kappa/2\pi<1$ GHz.  A decay rate of $1$ GHz is chosen as a benchmark value as it corresponds to a linewidth of a few ${\mu}$eV at these wavelengths, on par with the narrowest self-assembled InAs QD exciton linewidths that have been measured at cryogenic temperatures\cite{ref:Bayer}.  Indeed, because dissipation in a strongly-coupled QD-photon system can either be due to cavity decay or quantum dot dephasing, in Figure \ref{fig:sim_results_2} we examine the ratio of $g$ to the maximum decay rate in the system assuming a fixed QD dephasing rate $\gamma/2\pi$=1 GHz\footnote{Note that $\gamma \equiv \gamma_{\perp}$ is in general greater than half the total excitonic decay rate ($\gamma_{||}/2$) or radiative decay rate ($1/2\tau_{sp}$) for QD excitons, due to near-elastic scattering or dephasing events with, for example, acoustic phonons of the lattice.}.  This ratio is roughly representative of the number of coherent exchanges of energy (Rabi oscillations) that can take place between QD and photon.  We see that it peaks at a value of about 18 for a disk diameter $D\sim1.5$ $\mu$m.  For diameters smaller than this, loss is dominated by cavity decay due to radiation, while for larger diameters, the dominant loss mechanism is due to dephasing of the QD.  

For other types of atomic-like media besides the self-assembled InAs QDs considered here one need not assume a limit of $\gamma/2\pi=1$ GHz, and we note that due to the exponential dependence of $Q_{\text{rad}}$ and approximately linear dependence of $V_{\text{eff}}$ on microdisk diameter, $Q_{\text{rad}}/V_{\text{eff}}$ rapidly rises above $10^7$ for microdisks of diameter only $D=2.5$ $\mu$m.  These values of $Q_{\text{rad}}$ and $V_{\text{eff}}$ are comparable to those found in recent high-$Q$ photonic crystal microcavity designs\cite{ref:Srinivasan1,ref:Ryu5,ref:Song,ref:Kuramochi,ref:Zhang_Z}.  In fact a similar scaling for high-$Q$ planar photonic crystal microcavities, in which one may trade-off a linear increase in $V_{\text{eff}}$ for an exponential increase in $Q$, has recently been described by Englund, et al., in Ref. \cite{ref:Englund}.  For now, however, we take the ratio $g/\text{max}(\gamma,\kappa)$ with $\gamma/2\pi=1$ GHz as our metric, and as such focus on $1.5$-$2$ $\mu$m diameter microdisks.       

\section{Growth, Fabrication, and Test Set-up}
\label{sec:setup} 
The samples used were grown by molecular beam epitaxy, and consist of a single layer of InAs QDs embedded in an In$_{0.15}$Ga$_{0.85}$As quantum well, which is in turn sandwiched between layers of Al$_{0.30}$Ga$_{0.70}$As and GaAs to form a 255 nm thick waveguide layer.  This dot-in-a-well (DWELL) structure is grown on top of a 1.5 $\mu$m thick Al$_{0.70}$Ga$_{0.30}$As buffer layer that is later undercut to form the disk pedestal.  Growth parameters were adjusted\cite{ref:Stintz} to put the material's RT ground state emission peak at $\lambda=1317$ nm.  Fabrication of the microdisk cavities begins with the deposition of a $180$ nm thick Si$_{x}$N$_{y}$ etch mask layer through plasma-enhanced chemical vapor deposition.  This is followed by electron-beam lithography to define a linear array of disks, a post-development reflow of the resist, and a SF${_6}$/C$_{4}$F$_{8}$ inductively coupled plasma reaction ion etch (ICP-RIE) of the nitride mask.  The DWELL material is then etched using Ar/Cl$_{2}$ ICP-RIE, and the array of disks is isolated onto a mesa stripe through standard photolithography and an ICP-RIE etch.  Finally, the disks are undercut using a dilute solution of hydrofluoric acid (20:1 H$_2$O:HF); an image of a fabricated device is shown in Figure \ref{fig:SEM}.  During the fabrication, two goals were given special consideration.  As discussed within the context of silicon microdisks in Ref. \cite{ref:Borselli2}, elimination of radial variations in the disk geometry is important for reducing scattering loss; this is accomplished through optimization of the electron-beam lithography, and in particular, use of the post-development resist reflow technique of Ref. \cite{ref:Borselli2}.  In addition, a premium was placed on sidewall smoothness, even at the expense of sidewall verticality.  This required an optimization of the ICP-RIE processes used to etch both the Si$_{x}$N$_{y}$ mask and GaAs/AlGaAs waveguide layer.  In particular, a low bias voltage, C$_{4}$F$_{8}$-rich plasma is used to etch the Si$_{x}$N$_{y}$, and a low Cl$_{2}$ percentage is used in the Ar/Cl$_{2}$ etch of the waveguide to eliminate any sidewall pitting due to excessive chemical etching.          

The microdisks are studied in a photoluminescence (PL) measurement setup that provides normal incidence pumping and free-space collection from the samples. The pump laser is an $830$ nm laser diode that is operated continuous-wave, and the pump beam is shaped into a Gaussian-like profile by sending it through a section of single mode optical fiber, after which it is then focused onto the sample with an ultra-long working distance objective lens (NA $= 0.4$).  The free-space photoluminescence is first collected at normal incidence from the sample surface using the same objective lens for pump focusing, and is then coupled into a multi-mode fiber (MMF) using an objective lens with NA $= 0.14$. The luminescence collected by this MMF is wavelength resolved by a Hewlett Packard 70452B optical spectrum analyzer (OSA).   

The PL setup has been modified\cite{ref:Srinivasan11} to allow for devices to be probed by optical fiber tapers.  The fiber taper is formed by heating (with a hydrogen torch) and adiabatically stretching a single mode fiber until its minimum diameter is approximately $1$ $\mu$m.  It is mounted onto an acrylic mount that is attached to a motorized Z-axis stage (50 nm encoded resolution), so that the fiber taper can be precisely aligned to the microdisk, which is in turn mounted on a motorized XY stage.  When doing passive measurements of cavity $Q$, the taper input is connected to a scanning tunable laser (5 MHz linewidth) with a tuning range between $1420$-$1480$ nm, and the taper output is connected to a photodetector to monitor the transmitted power.  Alternately, when collecting emission from the microdisk through the fiber taper, the taper input is left unconnected and the output is sent into the OSA.

\section{Experimental Results}
\label{sec:results} 

We begin our measurements by using the fiber taper to passively probe the $Q$ of the microdisks.  Based on the simulations presented in Section {\ref{sec:sims}}, we have focused on $2$ $\mu$m diameter microdisks.  Due to the small diameter of these microdisks, the finite-element-calculated free-spectral range of resonant modes is relatively large, with resonances occurring at $1265$, $1346$, and $1438$ nm for the TE$_{p=1}$ WGMs with azimuthal mode numbers $m=11$,$10$, and $9$, respectively.  The simulations presented in Section {\ref{sec:sims}} were done for the TE$_{1,11}$ mode in the $\lambda=1200$ nm band due to the applicability of that wavelength region for future low temperature cQED experiments. However, for the current room-temperature measurements, the absorption due to the QD layer at those wavelengths is significant, so we probe the devices within the $\lambda=1400$ nm band ($\sim$100 nm red-detuned from the peak ground-state manifold QD emission).  At these longer wavelengths the radiation-limited $Q_{\text{rad}}$ for a given disk diameter will be smaller than its value in the shorter $\lambda=1200$ nm band.  Table \ref{table:FEMLAB_results} summarizes the properties of the TE$_{p=1}$ WGMs within the $1200$-$1400$ wavelength band for a $D=2$ $\mu$m microdisk with shape as shown in Fig. \ref{fig:SEM}.

\renewcommand{\arraystretch}{1.3}
\renewcommand{\extrarowheight}{0pt}
\begin{table}
\caption{Finite-element calculated TE$_{p=1,m}$ modes of a $D=2 \mu$m microdisk.}
\label{table:FEMLAB_results}
\begin{center}
\begin{tabularx}{0.84\linewidth}{YYYYY}
\hline
\hline
Mode label & $\lambda_{0}$ & $Q_{rad}$ & $V_{\text{eff}}$  & application \\
\hline
TE$_{1,9}$  & 1438 nm & $3.7{\times}10^5$ & 2.2 $(\lambda/n)^3$ & passive RT testing \\
TE$_{1,10}$ & 1346 nm & $1.9{\times}10^6$ & 2.5 $(\lambda/n)^3$ & RT lasers \\
TE$_{1,11}$ & 1265 nm & $9.8{\times}10^6$ & 2.8 $(\lambda/n)^3$ & low-T cQED \\
\hline
\hline
\end{tabularx}
\end{center}
\end{table}
\renewcommand{\arraystretch}{1.0}
\renewcommand{\extrarowheight}{0pt}

\begin{figure}[H]
\begin{center}
\epsfig{figure=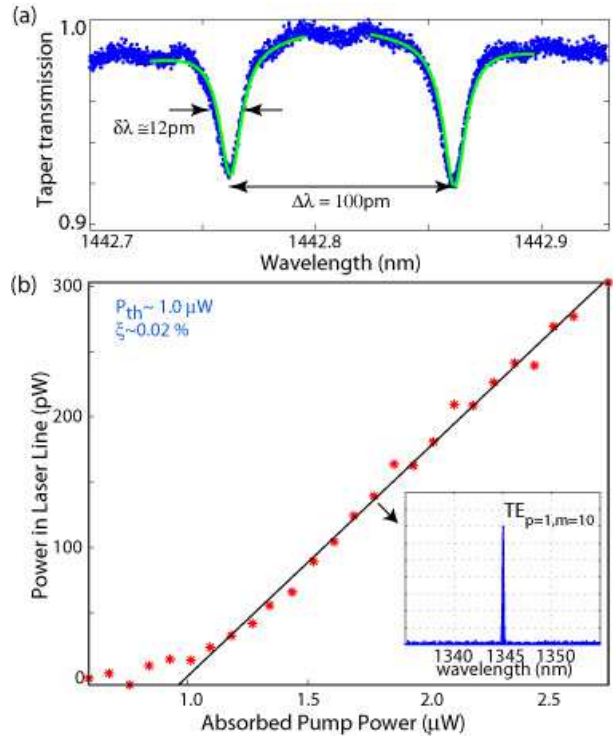, width=0.6\linewidth}
\caption{(a) Normalized transmission spectrum when a fiber taper is positioned a few hundred nm away from the microdisk edge. (b) Light-in-light-out (L-L) curve for a device operated with free-space collection.  The laser threshold absorbed pump power $P_{th}$ is $\sim$1.0 $\mu$W, and its differential efficiency $\xi\sim0.02\%$. (inset) Spectrum from the device above threshold, showing emission at $\lambda\sim1345$ nm corresponding to the TE$_{p=1,m=10}$ WGM.}
\label{fig:Q_plus_fs_coll}
\end{center}
\end{figure} 
    
Figure \ref{fig:Q_plus_fs_coll}(a) shows a wavelength scan of the transmitted signal when a fiber taper is positioned a few hundred nanometer away from the disk edge.  The doublet resonance appearing at $\lambda\sim1440$ nm in the spectrum is the signature of the standing wave modes described earlier\cite{ref:Weiss}.  The measured linewidths correspond to $Q$ factors of $1.2{\times}10^5$, and in general $Q$s of $0.9$-$1.3{\times}10^5$ have been measured for these $2$ $\mu$m diameter microdisks.  The $Q$s of these modes are approaching the radiation-limited value of $3.7{\times}10^5$, and are some of the highest measured values for near-IR wavelength-scale microcavities in AlGaAs\cite{ref:Yoshie3,ref:Gayral,ref:Srinivasan9,ref:Loffler}.  The corresponding cavity decay rates are $\kappa/2\pi\sim0.8-1.3$ GHz, over an order of magnitude smaller than the predicted coupling rate $g$ for an optimally placed QD.  In addition, these $Q$s, if replicated within the QD emission band at $\lambda=1300$ nm, are high enough to ensure that room-temperature lasing should be achievable from the single layer of QDs in these devices\cite{ref:Stintz}.   From calculations of the intrinsic radiation loss, the shorter $1300$ nm wavelength modes should in fact have a significantly increased $Q_{\text{rad}}$ of $1.9{\times}10^6$, although surface scattering may also slightly increase due to its approximate cubic dependence on wavelength\cite{ref:Borselli2}.   

\begin{figure}[t]
\begin{center}
\epsfig{figure=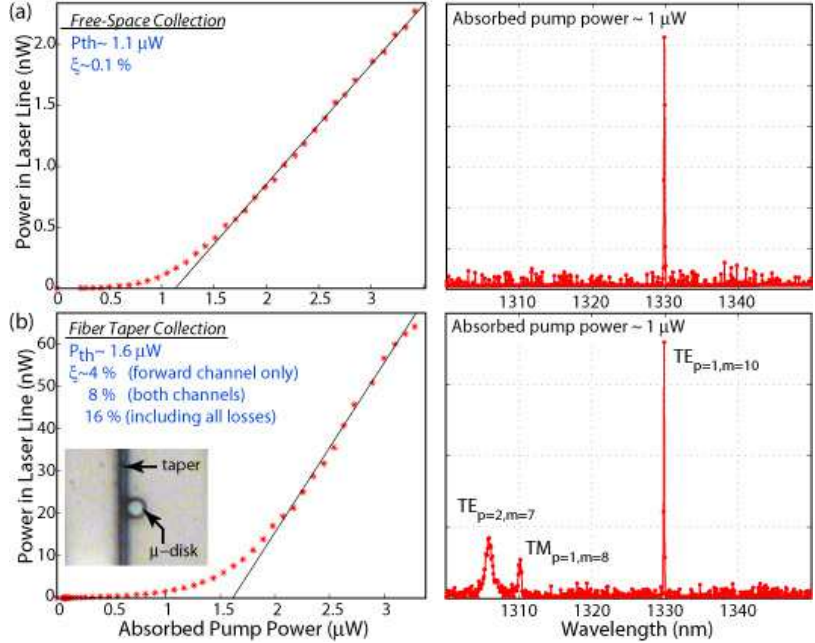, width=0.8\linewidth}
\caption{(a) (left) L-L curve for another microdisk device operated with free-space collection, with $P_{th}\sim1.1$ $\mu$W and $\xi\sim0.1\%$. (right) Spectrum from the device near laser threshold, showing emission at $\lambda\sim1330$ nm. (b) (left) L-L curve for the same device using an optical fiber taper to collect the emission.  $P_{th}$ has increased to 1.6 $\mu$W while $\xi$ increased to $4\%$ for collection in the forward fiber taper channel. (inset) Optical microscope image of the taper output coupler aligned to the microdisk. (right) Spectrum of the fiber taper collected light below threshold.}
\label{fig:fs_coll_vs_fiber_coll}
\end{center}
\end{figure}

The emission properties of the QD-containing microdisks are tested at room temperature by continuous-wave optical pumping through a high-NA objective lens at normal incidence and, initially, collecting the normal incidence emitted light through the same lens.  A light-in versus light-out (L-L) curve for one of the $D\sim2$ $\mu$m microdisks with a resonant emission peak at $\lambda\sim1345$ nm is shown in Figure \ref{fig:Q_plus_fs_coll}(b), and displays a lasing threshold kink at approximately $1.0$ $\mu$W of absorbed pump power.  The laser mode wavelength corresponds well with the TE$_{p=1,m=10}$ mode from finite-element simulations (see Table \ref{table:FEMLAB_results}).  The absorbed pump power is estimated to be $11\%$ of the incident pump power on the microdisk, and was determined assuming an absorption coefficient of $10^4$ cm$^{-1}$ for the GaAs layers and quantum well layer.  This threshold level is approximately two orders of magnitude smaller than those in recent demonstrations of RT, continuous-wave microdisk QD lasers\cite{ref:Ide2,ref:Yang_T2}, although the active regions in those devices contain five stacked layers of QDs while the devices presented here contain only a single layer of QDs.

The low lasing threshold of the device presented in Figure \ref{fig:Q_plus_fs_coll}(b) was consistently measured for the set of devices on this sample (approximately $20$ devices).  In Figure \ref{fig:fs_coll_vs_fiber_coll}(a) we show another L-L curve, this time for a device that has a TE$_{p=1,m=10}$ WGM emission peak at $\lambda=1330$ nm and has a threshold absorbed pump power of $1.1$ $\mu$W.  As demonstrated in Ref. \cite{ref:Srinivasan11}, the same fiber taper used to measure the cavity $Q$ can efficiently out-couple light from the lasing mode.  We do this by maintaining the free-space pumping used above while contacting a fiber taper to the side of the microdisk as shown in the inset of Figure \ref{fig:fs_coll_vs_fiber_coll}(b).  From the corresponding L-L curve (Fig. \ref{fig:fs_coll_vs_fiber_coll}(b)) we see that the laser threshold under fiber taper loading has increased from $1.1$ $\mu$W to $1.6$ $\mu$W, but in addition the differential laser efficiency $\xi$ is now $4\%$ compared to $0.1\%$ when employing free-space collection (Fig. \ref{fig:fs_coll_vs_fiber_coll}(a)-(b)).  Furthermore, because the microdisk modes are standing waves they radiate into both the forwards and backwards channels of the fiber.  With collection from both the forward and backward channels the differentially efficiency was measured to be twice that of the single forward channel.  Collecting from both channels and adjusting for all fiber losses in the system (roughly $50\%$ due to fiber splices and taper loss), the total differential laser efficiency with fiber taper collection is $16\%$.  Due to the difference in photon energy of the pump laser and microdisk emission, this laser differential efficiency corresponds to a conversion efficiency of $28\%$ from pump photons to fiber-collected microdisk laser photons.  $28\%$ is thus a \emph{lower} bound on the fiber-taper collection efficiency and/or quantum efficiency of the QD active region.   

In addition to the improved laser differential efficiency of the TE$_{p=1,m=10}$ laser mode when using the fiber taper to out-couple the laser light, we also see in the below-threshold spectrum of Figure \ref{fig:fs_coll_vs_fiber_coll}(b) that two additional resonances appear at $\lambda=1310$ nm and $\lambda=1306$ nm.  The long wavelength mode is identified as TM$_{p=1,m=8}$ and the short wavelength mode as TE$_{p=2,m=7}$ from finite-element simulations.  These modes are not discernible in the free-space collected spectrum due to their low radiation-limited $Q$ factors ($800$ and $5000$ for the TE$_{2,7}$ and TM$_{1,8}$, respectively), but show up in the taper coupled spectrum due to their alignment with the QD ground-state exciton emission peak and the heightened sensitivity of the taper coupling method.  The single-mode lasing and limited number of WGM resonances ($6$ when including the degeneracy of the WGMs) in the emission spectrum in these $D=2$ $\mu$m microdisks is a result of the large $80$-$100$ nm free-spectral-range of modes in the $1300$-$1500$ wavelength band.  As a result, one would expect the spontaneous emission factor ($\beta$) of these microdisk lasers to be relatively high.  

\begin{figure}[t]
\begin{center}
\epsfig{figure=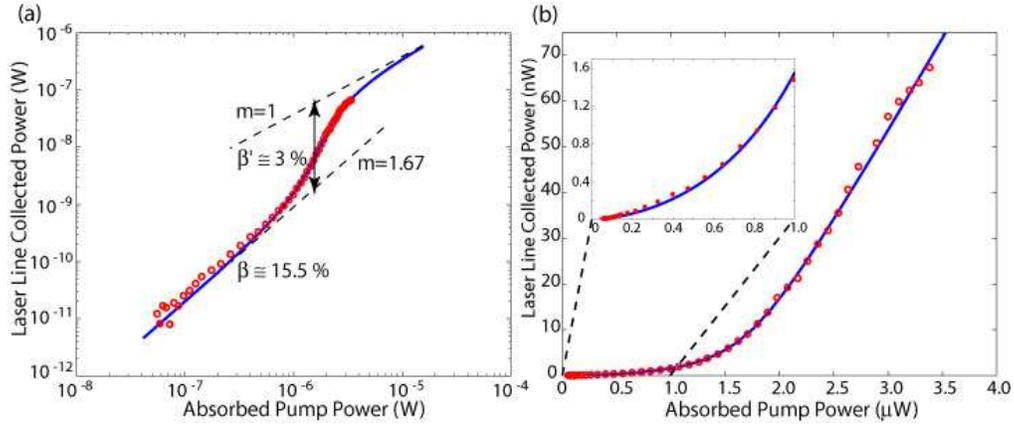, width=\linewidth}
\caption{L-L curve experimental data (red circles) and rate-equation model fit (blue line) to data for the fiber taper coupled laser of Figure \ref{fig:fs_coll_vs_fiber_coll}(b) : (a) log-log plot and (b) linear plot (inset shows deep sub-threshold data and fit). $\beta^{\prime}\sim3\%$ is the spontaneous emission factor estimated directly from the slope change in the data, and thus includes a large non-radiative component, while $\beta\sim15.5\%$ is the value used in the rate-equation model fit.}   
\label{fig:log_log}
\end{center}
\end{figure} 

A log-log plot of the fiber taper coupled laser emission of Figure \ref{fig:fs_coll_vs_fiber_coll}(b) is shown in Figure \ref{fig:log_log}(a) along with a rate-equation model fit to the data.  Of particular note is the well defined sub-threshold linear slope of the log-log plot.  In this case the sensitivity of the fiber taper collection allows for the sub-threshold slope to be accurately estimated at m=1.67, corresponding to a near quadratic dependence of spontaneous emission intensity on pump power (Figure \ref{fig:log_log}(b), inset) and indicating that there is likely significant non-radiative recombination.  Assuming that radiative recombination occurs as a bi-particle process\footnote{As has been discussed recently in Ref. \cite{ref:Pask} this may not be an accurate model for QD state-filling, but for our simple analysis here it will suffice.}, the larger than unity power law dependence of sub-threshold emission on pump power is indicative of single-particle non-radiative recombination processes such as surface recombination\cite{ref:Agrawal}.  Given the close proximity of the WGM laser mode to the periphery of the microdisk and the above-band pumping, the presence of significant surface recombination is not surprising.  Unfortunately, due to this large non-radiative component one can only provide a \emph{weak lower bound} $\beta^{\prime}$ for the $\beta$-factor directly from the L-L curve.  From Figure \ref{fig:log_log} we estimate $\beta\ge\beta^{\prime}\sim3\%$.  

A rate-equation model incorporating bi-particle spontaneous emission proportional to $N^2$ and surface recombination with a $N^{1.22}$ carrier dependence (the ratio of the power law dependences is set equal to the measured sub-threshold slope of m=1.67) is fit to the data and shown as a solid curve in Figure \ref{fig:log_log}.  In this model the measured fiber taper collection efficiency was used, along with the previously measured and estimated QD density, maximum gain, and quantum efficiency from stripe lasers\cite{ref:Stintz}.  An estimate for the actual radiative $\beta$-factor of $15.5\%$ was used, corresponding closely with the partitioning of spontaneous emission amongst the $6$ localized and high-$Q$ WGM resonances within the QD ground-state manifold emission band\footnote{This estimate was based upon considering Purcell enhancement at RT for QDs spatially and spectrally aligned with the WGMs ($F_{P}\sim6$), and suppression of spontaneous emission for QDs spatially and spectrally misaligned from the WGMs ($F_{P}\sim0.4$).  This simple estimate is consistent with accurate finite-difference time-domain calculations of similar sized microdisks\cite{ref:Vuckovic1}.}.  The reference spontaneous emission lifetime of the ground-state QD exciton in bulk was taken as $\tau_{sp}=1$ ns.  The data was fit by varying \emph{only} the effective surface recombination velocity.  As seen in Figure \ref{fig:log_log}, the fit is quite good over the entire sub-threshold and threshold regions of the laser data.  The inferred surface recombination velocity from the fit is $v_{s}\sim75$ cm/s, extremely slow for the AlGaAs material system\cite{ref:Coldren} but perhaps indicative of the fast capture rate of carriers, and consequent localization, into QDs\cite{ref:Sosnowski,ref:Yarotski}.  Due to the large perimeter-to-area ratio in these small $D=2$ $\mu$m microdisks, even with this low velocity the model predicts that laser threshold pump power is dominated by surface recombination with an effective lifetime $\tau_s\sim300$ ps.  Such a surface recombination lifetime has also been estimated by Baba, et al., in their recent work on QD-microdisk lasers\cite{ref:Ide}.             

The number of QDs contributing to lasing in these small microdisks can also be estimated.  From the finite-element simulations the area of the standing wave WGM lasing mode in the plane of the QD layer is approximately $1$ $\mu$m$^2$, and the predicted QD density for this sample is $300$ $\mu$m$^{-2}$, so that $\sim300$ QDs are spatially aligned with the cavity mode.  Assuming a RT homogeneous linewidth on the order of a few meV\cite{ref:Bayer}, compared to a measured inhomogeneous Gaussian broadening of $35$ meV, and considering the location of the lasing mode in the tail of the Gaussian distribution, we estimate $<10\%$ of these dots are spectrally aligned with the cavity mode.  By this estimate, on the order of $25$ QDs contribute to lasing.

\section{Conclusions} 
We have demonstrated fiber-coupled $2$ $\mu$m diameter quantum-dot-containing microdisks that have a quality factor $Q$ in excess of 10$^5$ for a predicted mode volume $V_{\text{eff}}$ as small as $\sim2.2(\lambda/n)^3$.  Such devices are predicted to be suitable for future experiments in single quantum dot, single photon experiments in cavity QED, where these $Q$ and $V_{\text{eff}}$ values can enable strong coupling at GHz-scale speeds.  An initial application of this work is in continuous wave, optically pumped microcavity lasers.  Here, the high $Q$ ensures that lasing can be achieved with the modest gain provided by a single layer of quantum dots, and combined with the ultra-small $V_{\text{eff}}$, results in thresholds as low as 1.0 $\mu$W of absorbed power.  In addition, the fiber taper coupling is shown to be an efficient method to collect the laser emission with a measured $28\%$ lower bound on out-coupling efficiency.     

This work was partly supported by the Charles Lee Powell Foundation.  The authors thank Christopher P. Michael, Paul E. Barclay, and Thomas J. Johnson for helpful discussions.  KS thanks the Hertz Foundation and MB thanks the Moore Foundation, NPSC, and HRL Laboratories for their graduate fellowship support.  


\begin{thebibliography}{1}

\bibitem{ref:Michler}
P. Michler, A. Kiraz, C. Becher, W. Schoenfeld, P. Petroff, L. Zhang, E. Hu,
  and A. Imamoglu, ``{A Quantum Dot Single-Photon Turnstile Device},'' Science
  {\bf 290,} 2282--2285 (2000).

\bibitem{ref:Santori}
C. Santori, M. Pelton, G. Solomon, Y. Dale, and Y. Yamamoto, ``{Triggered
  Single Photons from a Quantum Dot},'' Phys. Rev. Lett. {\bf 86,} 1502--1505
  (2001).

\bibitem{ref:Moreau}
E. Moreau, I. Robert, J. G\'{e}rard, I. Abram, L. Manin, and V. Thierry-Mieg,
  ``{Single-mode solid-state photon source based on isolated quantum dots in
  pillar microcavities},'' Appl. Phys. Lett. {\bf 79,} 2865--2867 (2001).

\bibitem{ref:Reithmaier}
J. Reithmaier, G. Sek, A. Loffer, C. Hoffman, S. Kuhn, S. Reitzenstein, L.
  Keldysh, V. Kulakovskii, T. Reinecke, and A. Forchel, ``{Strong coupling in a
  single quantum dot-semiconductor microcavity system},'' Nature {\bf 432,}
  197--200 (2004).

\bibitem{ref:Yoshie3}
T. Yoshie, A. Scherer, J. Hendrickson, G. Khitrova, H. Gibbs, G. Rupper, C.
  Ell, Q. Schenkin, and D. Deppe, ``{Vacuum Rabi splitting with a single
  quantum dot in a photonic crystal nanocavity},'' Nature {\bf 432,} 200--203
  (2004).

\bibitem{ref:Peter}
E. Peter, P. Senellart, D. Martrou, A. Lema$\hat{i}$tre, J. Hours, J.
  G\'{e}rard, and J. Bloch, ``{Exciton photon strong-coupling regime for a
  single quantum dot embedded in a microcavity},'' Phys. Rev. Lett. 95 (2005).

\bibitem{ref:Cao}
H. Cao, J. Xu, W. Xiang, Y. Ma, S.-H. Chang, S. Ho, and G. Solomon,
  ``{Optically pumped InAs quantum dot microdisk lasers},'' Appl. Phys. Lett.
  {\bf 76,} 3519--3521 (2000).

\bibitem{ref:Ide2}
T. Ide, T. Baba, J. Tatebayashi, S. Iwamoto, T. Nakaoka, and Y. Arakawa,
  ``{Room temperature continuous wave lasing InAs quantum-dot microdisks with
  air cladding},'' Opt. Express {\bf 13,} 1615--1620 (2005).

\bibitem{ref:Yang_T2}
T. Yang, O. Schekin, J. O'Brien, and D. Deppe, ``{Room temperature,
  continuous-wave lasing near 1300 nm in microdisks with quantum dot active
  regions},'' IEE Elec. Lett. 39 (2003).

\bibitem{ref:Kimble2}
H.~J. Kimble, ``{Strong Interactions of Single Atoms and Photons in Cavity
  QED},'' Physica Scripta {\bf T76,} 127--137 (1998).

\bibitem{ref:Cirac}
J. Cirac, P. Zoller, H. Kimble, and H. Mabuchi, ``{Quantum state transfer and
  entanglement distribution among distant nodes in a quantum network},'' Phys.
  Rev. Lett. {\bf 78,} 3221--3224 (1997).

\bibitem{ref:Knill}
E. Knill, R. Laflamme, and G. Milburn, ``{A scheme for efficient quantum
  computation with linear optics},'' Nature {\bf 409,} 46--52 (2001).

\bibitem{ref:Kiraz}
A. Kiraz, M. Atature, and A. Imamoglu, ``{Quantum-dot single-photon sources:
  Prospects for applications in linear optics quantum-information
  processing},'' Phys. Rev. A 69 (2004).

\bibitem{ref:McCall2}
S.~L. McCall, A.~F.~J. Levi, R.~E. Slusher, S.~J. Pearton, and R.~A. Logan,
  ``{Whispering-gallery mode lasers},'' Appl. Phys. Lett. {\bf 60,} 289--291
  (1992).

\bibitem{ref:Gayral}
B. Gayral, J.~M. G\'{e}rard, A. Lema$\hat{i}$tre, C. Dupuis, L. Manin, and
  J.~L. Pelouard, ``{High-$Q$ wet-etched GaAs microdisks containing InAs
  quantum boxes},'' Appl. Phys. Lett. {\bf 75,} 1908--1910 (1999).

\bibitem{ref:Srinivasan9}
K. Srinivasan, M. Borselli, T. Johnson, P. Barclay, O. Painter, A. Stintz, and
  S. Krishna, ``{Optical loss and lasing characteristics of high-quality-factor
  AlGaAs microdisk resonators with embedded quantum dots},'' Appl. Phys. Lett.
  {\bf 86,} 151106 (2005).

\bibitem{ref:Srinivasan11}
K. Srinivasan, A. Stintz, S. Krishna, and O. Painter, ``{Photoluminescence
  measurements of quantum-dot-containing semiconductor microdisk resonators
  using optical fiber taper waveguides},'' Phys. Rev. B {\bf 72,} 205318
  (2005).

\bibitem{ref:Spillane3}
S.~M. Spillane, T.~J. Kippenberg, K.~J. Vahala, K.~W. Goh, E. Wilcut, and H.~J.
  Kimble, ``{Ultrahigh-$Q$ toroidal microresonators for cavity quantum
  electrodynamics},'' Phys. Rev. A {\bf 71,} 013817 (2005).

\bibitem{ref:Borselli3}
M. Borselli, T. Johnson, and O. Painter,    (2005), manuscript in preparation.

\bibitem{ref:Andreani}
L. Andreani, G. Panzarini, and J.-M. G\'{e}rard, ``{Strong-coupling regime for
  quantum boxes in pillar microcavities:Theory},'' Phys. Rev. B {\bf 60,}
  13276--13279 (1999).

\bibitem{ref:Weiss}
D.~S. Weiss, V. Sandoghdar, J. Hare, V. Lef\`evre-Seguin, J.-M. Raimond, and S.
  Haroche, ``{Splitting of high-$Q$ Mie modes induced by light backscattering
  in silica microspheres},'' Opt. Lett. {\bf 20,} 1835--1837 (1995).

\bibitem{ref:Kippenberg}
T. Kippenberg, S. Spillane, and K. Vahala, ``{Modal coupling in traveling-wave
  resonators},'' Opt. Lett. {\bf 27,} 1669--1671 (2002).

\bibitem{ref:Borselli2}
M. Borselli, T. Johnson, and O. Painter, ``Beyond the Rayleigh scattering limit
  in high-Q silicon microdisks: theory and experiment,'' Opt. Express {\bf 13,}
  1515--1530 (2005).

\bibitem{ref:Bayer}
M. Bayer and A. Forchel, ``Temperature dependence of the exciton homogeneous
  linewidth in In$_{0.60}$Ga$_{0.40}$As/GaAs self-assembled quantum dots,''
  Phys. Rev. B {\bf 65,} 041308(R) (2002).

\bibitem{ref:Srinivasan1}
K. Srinivasan and O. Painter, ``{Momentum space design of high-Q photonic
  crystal optical cavities},'' Opt. Express {\bf 10,} 670--684 (2002).

\bibitem{ref:Ryu5}
H.-Y. Ryu, M. Notomi, and Y.-H. Lee, ``{High-quality-factor and
  small-mode-volume hexapole modes in photonic-crystal-slab nanocavities},''
  Appl. Phys. Lett. {\bf 83,} 4294--4296 (2003).

\bibitem{ref:Song}
B.-S. Song, S. Noda, T. Asano, and Y. Akahane, ``{Ultra-high-Q photonic
  double-heterostructure nanocavity},'' Nature Materials {\bf 4,} 207--210
  (2005).

\bibitem{ref:Kuramochi}
E. Kuramochi, M. Notomi, S. Mitsugi, A. Shinya, T. Tanabe, and T. Watanabe,
  ``{Photonic crystal nanocavity formed by local width modulation of
  line-defect with $Q$ of one million},'' In {\em LEOS 2005, Post-Deadline
  Session PD 1.1},   (IEEE Lasers and Electro-Optics Society, 2005).

\bibitem{ref:Zhang_Z}
Z. Zhang and M. Qiu, ``Small-volume waveguide-section high $Q$ microcavities in
  2D photonic crystal slabs,'' Optics Express {\bf 12,} 3988--3995 (2004).

\bibitem{ref:Englund}
D. Englund, I. Fushman, and J. Vu\v{c}kovi\'{c}, ``General recipe for designing
  photonic crystal cavities,'' Optics Express {\bf 13,} 5961--5975 (2005).

\bibitem{ref:Stintz}
A. Stintz, G. Liu, H. Li, L. Lester, and K. Malloy, ``Low-Threshold Current
  Density 1.3-$\mu$m InAs Quantum-Dot Lasers with the Dots-in-a-Well (DWELL)
  structure,'' IEEE Photonics Tech. Lett. {\bf 12,} 591--593 (2000).

\bibitem{ref:Loffler}
A. Loffler, J. Reithmaier, G. Sek, C. Hofmann, S. Reitzenstein, M. Kamp, and A.
  Forchel, ``{Semiconductor quantum dot microcavity pillars with high-quality
  factors and enlarged dot dimensions},'' Appl. Phys. Lett. {\bf 86,} 111105
  (2005).

\bibitem{ref:Pask}
H. Pask, H. Summer, and P. Blood, ``{Localized Recombination and Gain in
  Quantum Dots},'' In {\em Tech. Dig. Conf. on Lasers and Electro-Optics,
  CThH3},   (Optical Society of America, Baltimore, MD, 2005).

\bibitem{ref:Agrawal}
G.~P. Agrawal and N.~K. Dutta, {\em {Semiconductor Lasers}} (Van Nostrand
  Reinhold, New York, NY, 1993).

\bibitem{ref:Vuckovic1}
J. Vu\v{c}kovi\'{c}, O. Painter, Y. Xu, A. Yariv, and A. Scherer, ``{FDTD
  Calculation of the Spontaneous Emission Coupling Factor in Optical
  Microcavities},'' IEEE J. Quan. Elec. {\bf 35,} 1168--1175 (1999).

\bibitem{ref:Coldren}
L.~A. Coldren and S.~W. Corzine, {\em {Diode Lasers and Photonic Integrated
  Circuits}} (John Wiley \& Sons, Inc., New York, NY, 1995).

\bibitem{ref:Sosnowski}
T. Sosnowski, T. Norris, H. Jiang, J. Singh, K. Kamath, and P. Bhattacharya,
  ``{Rapid carrier relaxation in In${0.4}$Ga$_{0.60}$As/GaAs quantum dots
  characterized by differential transmission spectroscopy},'' Phys. Rev. B {\bf
  57,} R9423--R9426 (1998).

\bibitem{ref:Yarotski}
D. Yarotski, R. Averitt, N. Negre, S. Crooker, A. Taylor, G. Donati, A. Stintz,
  L. Lester, and K. Malloy, ``{Ultrafast carrier-relaxation dynamics in
  self-assembled InAs/GaAs quantum dots},'' J. Opt. Soc. Am. B {\bf 19,}
  1480--1484 (2002).

\bibitem{ref:Ide}
T. Ide, T. Baba, J. Tatebayashi, S. Iwamoto, T. Nakaoka, and Y. Arakawa,
  ``{Lasing characteristics of InAs quantum-dot microdisk from 3K to room
  temperature},'' Appl. Phys. Lett. {\bf 85,} 1326--1328 (2004).

\end{thebibliography}

\end{document}